\documentclass[usenatbib]{mn2e}

\usepackage[psamsfonts]{amssymb}
\usepackage[dvips]{graphicx}

\def\etal{{\it et al.}}

\def\msun{{\rm ~M$_{\odot}$}}
\def\rsun{{\rm ~R$_{\odot}$}}
\def\ie{{\rm i.e.}}

\def\aj{{\it Aj}}
\def\apj{{\it ApJ}}
\def\mnras{{\it MNRAS}}
\def\apjl{{\it ApJL}}

\def\araa{{\it ARA\&A}}

\def\aap{{\it A\&A}}

\title[Blue Stragglers as Stellar Collision Products]{Blue Stragglers as Stellar Collision Products: the Angular Momentum Question}
\author[Alison Sills, Tim Adams and Melvyn B. Davies]
{Alison Sills$^{1}$, Tim Adams$^{2}$ and Melvyn B.~Davies$^{3,2}$ \\
$^{1}$Department of Physics and Astronomy, McMaster University, Hamilton, Ontario, Canada L8S 4M1 \\
$^{2}$Department of Physics and Astronomy, University of Leicester, Leicester, LE1 7RH\\
$^{3}$Lund Observatory, Box 43, SE-221 00, Lund, Sweden}

\date{Received ** *** 2004; in original form 2004 *** **}

\begin{document}

\maketitle

\begin{abstract}
We investigate the structure and evolution of blue stragglers stars
which were formed from direct stellar collisions between main sequence
stars in globular clusters. In particular, we look at the rotational
evolution of the products of off-axis collisions. As found in previous
work, such blue stragglers initially have too high an angular momentum
to contract down to the main--sequence. We consider angular momentum
loss through either disc locking or locking to an outflowing wind and
show that both methods allow the merged object to shed sufficient
angular momentum to contract down to the main-sequence. 
\end{abstract}

\begin {keywords}
convection -- hydrodynamics -- blue stragglers -- stars: rotation
\end{keywords}

\section{Introduction and Motivation}

Blue stragglers are those main sequence stars which are bluer and
brighter than the main sequence turnoff in clusters, and must
therefore be younger than the bulk of the cluster population. In
globular clusters especially, which are devoid of an interstellar
medium from which to form young stars, two viable mechanisms for blue
straggler formation exist. 

First, blue stragglers could be a result of a collision between two
main sequence stars. Globular clusters are crowded places, with
densities in post-core-collapse clusters in excess of 10$^5$
stars/pc$^3$. Within such a crowded environment it is inevitable that
collisions will occur between stars. Indeed collisions have been
invoked as a possible mechanism for explaining the apparent paucity of
red giants in the cores of globular clusters \citep*{ADS04}.  It seems
reasonable that collisions could play a role in the creation of blue
stragglers.

\begin{figure}
\includegraphics[clip,width=0.95\linewidth]{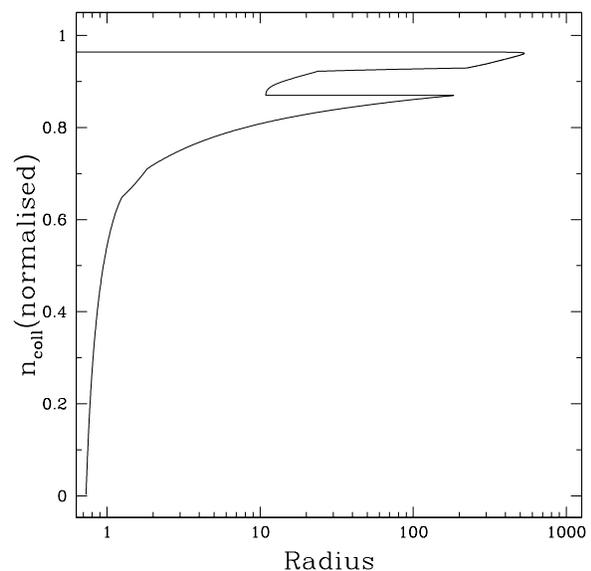}
\caption{The normalised collision probability for the 0.4~\msun model
used in this investigation as a function of its radius (which is a
tracer of its age). As can be seen the star is most likely to undergo
a collision while it is on the main-sequence. The paths for the 0.6
and 0.8\msun models are very similar to this.}
\label{zigzag}
\end{figure}

In Fig.~\ref{zigzag} We plot the normalised collision probability for
the 0.4\msun model used throughout this work \citep{BD99}. The other
stellar models follow a similar path. As can be seen in this figure,
the vast majority of collisions are likely to occur for a star whilst
it is on the main-sequence. This is simply because the main-sequence
lifetimes of these low mass stars is very long.

Alternatively, blue stragglers may be formed within primordial
binaries. \citet{PS00} describe observations of blue stragglers
observed in the field. They conclude that they must be produced by a
period of mass transfer within a binary system. Mass from the more
evolved star can pass to the less evolved companion as it expands and
fills its Roche lobe.  The impact of such primordial systems on the
entire blue straggler population within a globular cluster is
discussed in \citet{DPd04}.

We currently do not have good diagnostics to discriminate between
collisional blue stragglers and binary mergers observationally. A
number of groups have modelled direct physical collisions between main
sequence stars \citep{BH87,LRS96,SBH97,SFLRW01} and the subsequent
evolution of the collision products \citep{SLBDRS97,SFLRW01} to
investigate positions in the colour-magnitude diagram, rotation rates
and surface abundances. The equivalent work has not yet been done for
binary mergers. This is mostly because of the difficulty is following
the evolution of a hydrodynamic system (the mass transfer between
binary components) for the length of time required for the system to
merge. This timescale is unknown but certainly large, on the order of
half a billion years \citep[e.g.][]{R81}.

The details of the evolution of blue stragglers from either formation
mechanism is both interesting in its own right, and necessary for
detailed studies of the dynamics of globular clusters. Over the last
decade, it has become more obvious that dynamical models of globular
clusters needed to incorporate more realistic stellar effects. Rather
than treating the cluster as a self-gravitating system of equal-mass
point masses, the stars in the cluster were allowed to have a spectrum
of masses. This required some discussion of stellar evolution, so that
high mass stars were not still influencing the dynamics of the cluster
well after they had evolved off the main sequence. Binary stars,
particularly primordial binaries, became critically important for
halting core collapse in clusters, necessitating some form of binary
evolution in the dynamics codes.  Stars were also given finite radii,
which allowed the study of stellar collisions. All these new levels of
complexity point in one direction -- understanding the interplay
between stellar dynamics and stellar evolution is an essential
component in understanding globular clusters and their populations. 

Stellar collisions are one component of this problem, and fortunately,
the products of some stellar collisions are readily observed in the
colour-magnitude diagram as blue stragglers. Therefore, by
understanding the current population of blue stragglers in clusters,
we can begin to probe the past dynamical history of the cluster. This
is particularly true if we limit ourselves to regions of the cluster
where collisions are likely to dominate (i.e. the centres of dense
clusters), which has been possible since the advent of the Hubble
Space Telescope. We have made some comparisons between HST
observations and theoretical predictions in a number of clusters
\citep{SB99, SFLRW01, FSRPB03}. In order to make these
comparisons, we need to understand the evolution of blue stragglers in
the colour magnitude diagram over their entire lifetimes.

In previous work, we used the results of smoothed particle
hydrodynamics (SPH) simulations of stellar collisions as starting
models of stellar evolution calculations, and we produced detailed
evolutionary models for a variety of collision products relevant to
globular clusters. For head-on collisions at both medium and high SPH
resolution \citep{SLBDRS97,SADB02}, these models were
very successful, and we understand the evolution of these collision
products quite well. However, the limiting case of exactly head-on
collisions is unrealistic. Most collisions will have some non-zero
impact parameter. These off-axis collision products have substantial
angular momentum from their initial configuration (even assuming
initially non-rotating stars). Even a collision with a very small, but
non-zero impact parameter can result in a collision product with
substantial angular momentum. This poses a problem for the subsequent
evolution of the blue straggler.

It has been known for some time that immediately after the collision,
the stars are rotating quite rapidly \citep{LRS96}. They have total
angular momenta up to 10 times larger than low-mass pre-main sequence
stars ($\sim 10^{51}$ g cm$^2$ s${-1}$ for the collision products),
and, like pre-main sequence stars, have very large radii. Therefore,
they are rotating rapidly, but not unduly so. Like pre-main sequence
stars, the collision products evolve by contracting towards the main
sequence. Unlike pre-main sequence stars, however, collision products
do not have surface convection zones, and therefore cannot lose
angular momentum via a magnetic wind
\citep*{CDP95}. The contraction of a pre-main sequence star combined
with this wind loss results in a normal main sequence star with
approximately 1\% of its initial angular momentum. The contraction of
a collision product without this wind loss results in a star which is
rotating at greater than its break-up velocity long before it reaches
the main sequence. The conclusion of \citet{SFLRW01} was that either
blue stragglers cannot be created by stellar collisions because they
tear themselves apart through rapid rotation, or that some other
mechanism for removing angular momentum must occur.

Very little observational data exists for rotation rates of blue
stragglers. The only blue stragglers in a globular cluster to have its
rotation rate measured are BSS 19 in 47 Tucanae \citep{SSL97}, which
has $v \sin i = 155 \pm 55 $ km s$^{-1}$; and M3-17 \citep{DLOZS04}
with $v \sin i = 200 \pm 50$ km s$^{-1}$. There is also an upper limit
of $v \sin i = 50$ km s$^{-1}$ on NGC6752-11 \citep{DLOZS04}. The
other blue stragglers with measured rotation rates are found in the
old open cluster M67
\citep{PCL84}. These stars have $v \sin i$ between 10 and 120 km s$^{-1}$,
which is somewhat lower than most stars of the same temperature in
younger clusters. Most of the blue stragglers in M67 are found in
binary systems, which may indicate that their formation mechanism is
different from those in globular clusters. The lack of observation of
rotation velocities for globular cluster blue stragglers means that we
cannot directly compare our predictions of rotation rates with
data. Instead, we are restricted to looking at the effect of rotation
on the position of stars in the colour-magnitude diagram, and making
inferences about what is plausible.

Very recently, some interesting observational results have been
published which may shed new light on this problem. \citet{DLOZS04}
have discovered very sparse circumstellar discs around six blue
stragglers in globular clusters, out of the 55 or so for which they
have HST/STIS spectra. Since magnetic locking of the star to a disc is
the leading mechanisms for angular momentum loss in these systems,
this result is encouraging. Four of the six blue stragglers have no
measure rotation rates. In this small sample, there is not enough
information to place strong constraints on models yet.

In this paper, we revisit the question of structure and evolution of
collision products. In particular we restrict ourselves to off-axis
collisions between main sequence stars relevant to globular
clusters. We concentrate on investigating a plausible mechanism for
removing angular momentum from the collision products early in their
post-collision evolution.  Section \ref{SPH} describes the simulations of the
collisions themselves, while section 3 outlines our approach to the
evolution of the collision products and how the angular momentum
evolution is treated. Section 4 draws conclusions based on these
calculations and discusses how these results may impact our
understanding of blue straggler populations in globular clusters.

\section{Smoothed Particle Hydrodynamics Simulations \label{SPH}}

\subsection{Method}

The simulations of stellar collisions discussed in this paper are very
similar to those presented in \citet{SADB02}. We used the smoothed
particle hydrodynamics technique \citep{B90,M92}. Our
three-dimensional code uses a tree to solve for the gravitational
forces and to find the nearest neighbours \citep{B90}. We use the
standard form of artificial viscosity with $\alpha = 1$ and $\beta =
2.5$ \citep{M92}, and an adiabatic equation of state. The
thermodynamic quantities are evolved by following the change in
internal energy. Both the smoothing length and the numbers of
neighbours can change in time and space. The smoothing length is
varied to keep the number of neighbours approximately constant ($\sim
50$). The code is the parallel version of the code described in
\citet*{BBP95} It was parallelized using OpenMP, and was run on the
SGI Origin 3800 operated by the UK Astrophysical Fluids Facility
(UKAFF) based at the University of Leicester, and on the SHARCNet
facilities at McMaster University.

We modelled collisions between main sequence stars of masses 0.4, 0.6
and 0.8 \msun with impact parameters between 0.25 and 1 in units of
the sum of the parent star radii, as outlined in Table
\ref{SPHsetup}. The initial stellar models were calculated using the
Yale stellar evolution code (YREC, \citep{GDPK92}) and had a
metallicity of Z=0.001 and an age of 13.5 Gyr. This age was chosen so
that the 0.8 \msun star was just at the turnoff. The particles were
initially distributed on an equally spaced grid, and their masses were
varied until the density profile matched that of the stellar
model. The outermost SPH particles were set so that they were within
one smoothing length of the outer radius of the star as calculated by
YREC. This positioning of the particles and the variable particle
masses were chosen in order to resolve the majority of the star,
including particularly the outer layers. The stars were given a
relative velocity at infinity of $v_{\infty}$=10 km s$^{-1}$, which is
a reasonable value for globular cluster stars. They were set up on
almost parabolic orbits with a pericentre separation as shown in Table
1. The stars are initially non-rotating and separated by 5 \rsun. The
centre of mass of the system is set to be at the origin.

\begin{table}
\centering
\begin{minipage}{140mm}
\caption{Description of SPH simulations performed \label{SPHsetup}}
\begin{tabular}{lllllll}
\hline
Run name & m1 & m2 & rp & Npart & r1 & r2 \\
 & \msun & \msun & r1+r2 & &\rsun &\rsun \\
\hline
M44Q & 0.4 & 0.4 & 0.25 &  69 464 & 0.40 & 0.40 \\
M44H & 0.4 & 0.4 & 0.5  &  86 784 & 0.40 & 0.40 \\
M44F & 0.4 & 0.4 & 1.0  & 100 994 & 0.40 & 0.40 \\
M46Q & 0.4 & 0.6 & 0.25 &  44 702 & 0.40 & 0.61 \\
M46H & 0.4 & 0.6 & 0.5  &  36 750 & 0.40 & 0.61 \\
M48Q & 0.4 & 0.8 & 0.25 & 100 514 & 0.40 & 1.15 \\
M66Q & 0.6 & 0.6 & 0.25 & 100 034 & 0.61 & 0.61 \\
M66H & 0.6 & 0.6 & 0.5  & 100 034 & 0.61 & 0.61 \\
M66F & 0.6 & 0.6 & 1.0  & 100 034 & 0.61 & 0.61 \\
M68Q & 0.6 & 0.8 & 0.25 & 100 034 & 0.61 & 1.15 \\
M88Q & 0.8 & 0.8 & 0.25 & 100 034 & 1.15 & 1.15 \\
\hline
\end{tabular}
\end{minipage}
\end{table}

\subsection{Results}

The global properties of our collision products agree with those
presented in \citet{SFLRW01} for those simulations with the same
initial conditions. In Table \ref{SPHresults} we give some information
about the final state of each collision. In column 2 we give the total
angular momentum of the collision product. In column 3 we give the
ratio of the rotational energy about the z axis (T) to the potential
energy (W) of the system. A rotating system is considered to be
secularly unstable if $T/\vert W \vert \geq 0.14$, and dynamically
unstable if $T/\vert W \vert \geq 0.26$ \citep{T00}. All our collision
products begin their lives considerably below these values. In column
4 we give the mass of the collision product immediately after the
collision. For a given choice of parent masses, the mass loss due to
the collision is larger for smaller impact parameters.

\begin{figure}
\includegraphics[clip,width=0.95\linewidth]{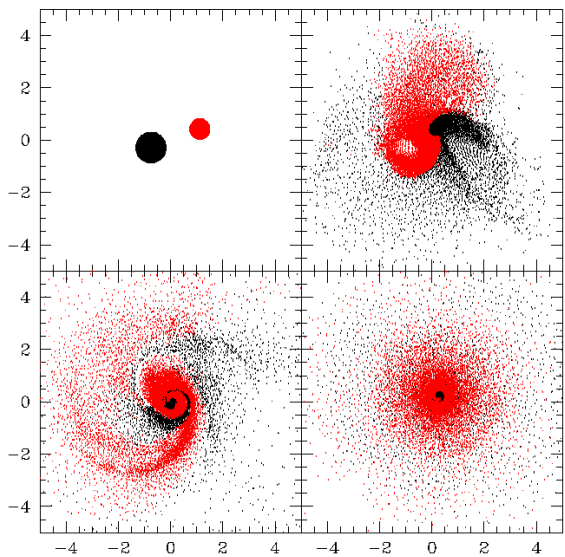}
\caption{A time series plot of those particles contained within 2$h$ of
the z=0 plane. This is a collision between a 0.6 (black particles) and
and a 0.4\msun (red particles) star, with a minimum distance of
approach of 0.255\rsun~(\ie 0.25 $R_1+R_2$). As can be seen, it is the
material from the heavier star which ends up at centre of the merged
object. It is the composition of the core which will determine how
long a particular blue straggler will spend on the main sequence
compared to a star of the same mass produced through the collapse of
the initial gas cloud.}
\label{blue:partplot}
\end{figure}

In Fig.~\ref{blue:partplot} we show a time series plot from a
collision between a 0.6 and a 0.4 \msun~star with a minimum distance
of approach of 0.25(R$_1$+R$_2$).  As was found in previous work, it
is the material with the highest entropy (in most cases, this is the
material from the most evolved star) which ends up at the centre of
the blue straggler. However, there is some mixing of material and some
of the material from the less evolved star does make it into the core
of the collision product. A similar pattern is found in all of the
collisions.

\begin{figure}
\includegraphics[clip,width=0.95\linewidth]{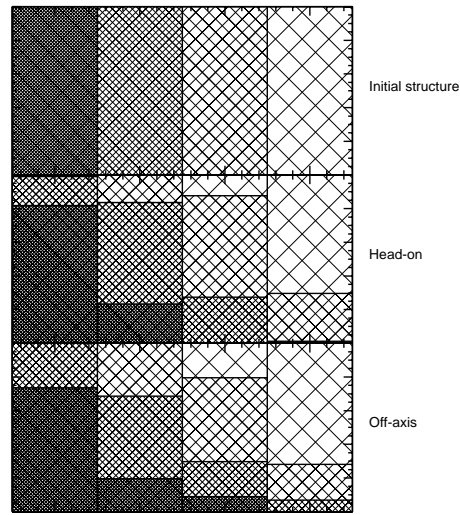}
\caption{A plot of where the mass now in the collision product discussed above
originally came from. The four columns in each plot represent four
equally space mass bins. The darkest hatching represents material that
came from the inner quarter of the parents stars, whilst the widest
hatching represents material which came from the outer quarter of the
parents stars. Shown are two blue straggler models, a head-on
collision and an off-axis collision. As can be seen, the innermost
quarter of the blue straggler is composed primarily of processed
material from the parent stars. This indicates that the blue straggler
will have a shorter lifespan than a main sequence star of equivalent
mass.}
\label{hatched}
\end{figure}

This is again present in Fig.~\ref{hatched}. Here we have plotted a
histogram similar to that shown in \citet{BH87}. The different levels
of shading represent where material has come from within the star. All
the stars are split into four bins based on radius from their
centres. Where the material ends up in the bins of the blue straggler
is then shown in the bottom histogram. The heaviest shading represents
the material from the inner quartile of the parent stars, whilst the
lightest shading represents the outermost quartile. Again we can
clearly see that the inner regions of the blue straggler are made u
the material from the innermost regions of the parent stars, and thus
the blue straggler is not well mixed. This is most true of the head-on
collision, but still quite apparent in the offaxis collision.

\begin{table}
\centering
\begin{minipage}{140mm}
\caption{Results of SPH simulations.  \label{SPHresults}}
\begin{tabular}{llll}
\hline
Run name & J & T/$\vert W \vert$ & $M_{\rm final}$\\
 & g cm$^2$ s$^{-1}$ &  & \msun \\
\hline
M44Q & $6.96 \times 10^{50}$& 0.038 & 0.748 \\
M44H & $9.85 \times 10^{50}$& 0.067 & 0.772 \\
M44F & $9.89 \times 10^{50}$& 0.069 & 0.796 \\
M46Q & $1.03 \times 10^{51}$& 0.033 & 0.856 \\
M46H & $1.48 \times 10^{51}$& 0.039 & 0.836 \\
M48Q & $1.58 \times 10^{51}$& 0.001 & 1.06  \\
M66Q & $1.11 \times 10^{51}$& 0.026 & 1.16 \\
M66H & $1.57 \times 10^{51}$& 0.047 & 1.17 \\
M66F & $2.22 \times 10^{51}$& 0.052 & 1.19 \\
M68Q & $2.32 \times 10^{51}$& 0.038 & 1.37 \\
M88Q & $3.34 \times 10^{51}$& 0.008 & 1.48 \\
\hline
\end{tabular}
\end{minipage}
\end{table}

\begin{figure}
\includegraphics[clip,width=0.95\linewidth]{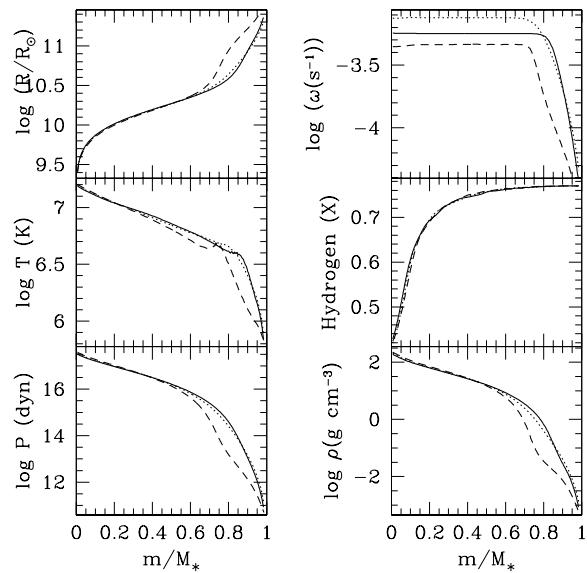}
\caption{Structure of off-axis collision product for a collision between 
two 0.6 \msun stars, with an impact parameter of 0.25 (solid line), 
0.5 (dotted line) and 1.0 (dashed line). \label{profiles}}
\end{figure}

Figure \ref{profiles} shows the structure of off-axis collision
products between two 0.6 \msun stars with a variety of impact
parameters. The biggest difference between the three different
collision products is their internal rotation rates. For the most
part, their structural parameters are very similar.

\begin{figure}
\includegraphics[clip,width=0.95\linewidth]{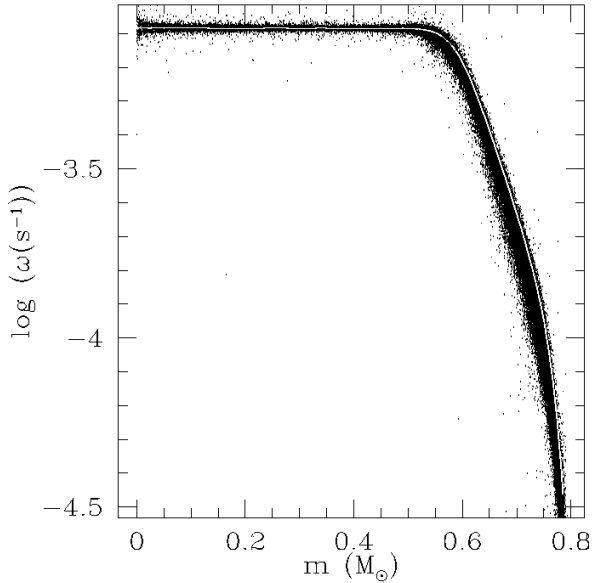}
\caption{Angular velocity as a function of mass fraction for a collision 
between two 0.4 \msun stars, with an impact parameter of 1.0. The dots
show the results for each SPH particle, and the solid line is the
average. \label{Wprofile}}
\end{figure}

Figure \ref{Wprofile} gives the angular velocity of each SPH particle
in the M44F collision as a function of mass fraction. There are two
important things to be noticed in this figure. First, the SPH
particles form a very tight sequence, indicating that when we
transform from our inherently 3-D SPH simulation to our inherently 1-D
stellar evolution calculation, we have correctly identified the axis
of rotation of the collision product, and that the collision product
is rotating in a coherent manner. Secondly, the angular velocity of
this collision product (and all the products produced in these
simulations) is almost constant for a large fraction of the interior
of the star. While similar simulations show a reasonably flat omega
profile \citep{SFLRW01}, the profiles derived from these simulations
are even flatter. This is probably because of our use the the Monaghan
artificial viscosity, which is known to introduce some shear viscosity
that will artificially flatten rotation profiles.

We do not see evidence for the presence of accretion discs surrounding
the collision products at the end of our simulations for any
combination of parent masses and impact parameter. This is in
disagreement with the findings of \citet{BH87}. They collided two
equal-mass n=1.5 polytropes with a variety of impact parameters and
velocities, and found that discs form in many of their collisions. The
collisions which correspond most closely to the globular cluster
collisions we are modelling have $v_{\infty}=0$, and they give disc
masses between 3 and 10\% of the mass of system for impact parameters
between 0.31 and 0.953 $R_1+R_2$, where $R_1$ and $R_2$ are the radii
of the impacting stars.

The difference between our results and those of \cite{BH87} may be
related to a problem of definition. Most of our off axis simulations
result in stars which are significantly flattened by rotation, and
have rather oblate density contours at large distances from the star
flattened in the z direction. (The stars are rotating about the z
axis). However, none of the density contours have an axis ratio which
is extreme enough to be called a disc. In fact, these results look
similar in content to figure 9 of \cite{BH87}, in which their
accretion disc is approximately as thick as the (flattened) radius of
the central star.  It is also possible that the presence of a disc
might have been a numerical artifact caused by the lower resolution of
the earlier simulations. Although we haven't seen direct evidence for
a disc around the merged object, there is some material which is bound
to the central object, although not apparently part of the blue
straggler. Perhaps this material could eventually go on to form a disc
around the central object.


\section[Evolution]{Evolution of collision products}\label{blue:ev}

With the SPH simulations completed, the next task is to perform the
evolution of the models. Our SPH models showed, as in previous work,
that they had too much angular momentum to contract down to the
main-sequence. If these collision products are going to form some
portion of the blue straggler population seen in globular clusters,
they need a method of losing some of this angular momentum.

\subsection{Mass and angular momentum loss}

Let us consider what is likely to happen to the blue straggler as it
contracts down toward the main-sequence. As the star contracts it has
to conserve angular momentum, so material is forced to rotate about
the rotation axis faster. As this happens the star becomes deformed,
becoming somewhat oblate. It is here that a new mechanism might be
able to take effect. We know that the star can contract down to the
point where it reaches its break up speed. At this point we could get
mass loss as surface material becomes unbound from the star.  This
mass loss could do one of two things: it could take angular momentum
away to infinity, thus reducing the angular momentum of the merged
object, or alternatively it might form a disc around the star. If the
star then has a strong magnetic field (which seems reasonable for most
of the stars produced in the collisions detailed here as they are
rapidly rotating) then it is possible for the disc and star to become
locked, thus spinning down the star. Or the same magnetic field could
be locked to the freely flowing material thus providing a torque in
the same way as a stellar wind.

We need to examine how these two effects will alter the spin of the
star if we are to incorporate their effects in the stellar evolution
code. 

\subsubsection{Disc locking}

Let us first look at the effects of disc locking. The prescription
that we have used follows on from the work of \citet{AC96}. The torque
on a star from an annulus of a magnetically locked accretion disc, of
width $\Delta R$ is related to the transfer of angular momentum
through magnetic stresses:

\begin{eqnarray}
\Gamma_{\rm annulus}=B_\phi B_{\rm z} R^2\Delta R
\label{blue:torque}
\end{eqnarray}

\noindent where $B_{\rm z}$ is the vertical component of the magnetic
field at a distance $R$ from the star and $B_{\phi}$is the toroidal
component of the magnetic field. These two magnetic components may be
expressed as:

\begin{eqnarray}
B_{\rm z}=\frac{\mu}{R^3}
\label{blue:bz}
\end{eqnarray}

\noindent where $\mu$ is the stellar dipole moment. \citet{LP92} give the toroidal component as:

\begin{eqnarray}
B_\phi=B_{\rm z}\left(\frac{\Omega(R)-\Omega_\star}{\Omega(R)}\right)
\label{blue:bphi}
\end{eqnarray}

\noindent where $\Omega$ is the angular velocity of the Keplerian disc and
$\Omega_\star$ is the angular velocity of the star\footnote{This
relation breaks down near the co-rotation radius, at this point
$B_{\phi} \approx B_{\rm z}$}.

The torque on the star will produce a change in its angular momentum,
thus:

\begin{eqnarray}
\frac{d J}{dt}&=&\Gamma_{\rm annulus}\nonumber \\
&=&\frac{\mu^2}{R^4}\left(\frac{\Omega(R)-\Omega_\star}{\Omega(R)}\right)
\Delta R 
\end{eqnarray}

\noindent and a change in the angular momentum of the body will alter
its rotation rate:

\begin{eqnarray}
\frac{d
J}{dt}=\Omega_\star\frac{dI_\star}{dt}+I_\star\frac{d\Omega_\star}{dt}
\end{eqnarray}

\begin{eqnarray}
\frac{d\Omega_\star}{dt}=\frac{1}{I_\star}\left(\Gamma_{\rm annulus}-
\Omega_\star\frac{dI_\star}{dt}\right) 
\label{blue:domega}
\end{eqnarray}

To get a proper estimate of the torque from the entire disc we have to
integrate Eqn.~\ref{blue:torque} over the extent of the disc using
the definitions of Eqn.s~\ref{blue:bz} and~\ref{blue:bphi}. If
we assume that it extends from some inner radius $R_{\rm in}$ out to
infinity then we get:

\begin{eqnarray}
\Gamma_{\rm disc}=\frac{\mu^2}{3}\left(\frac{1}{R_{\rm in}^3}-\frac{2
\Omega_\star}{\left(G M_\star R_{\rm in}^3\right)^{\frac{1}{2}}}\right)
\end{eqnarray}

\noindent Thus the spin down rate of the star is now given by:

\begin{eqnarray}
\frac{d\Omega_\star}{dt}=\frac{1}{I_\star}\left[
\frac{\mu^2}{3}\left(\frac{1}{R_{\rm in}^3}-\frac{2 
\Omega_\star}{\left(G M_\star R_{\rm in}^3\right)^{\frac{1}{2}}}\right)- 
\Omega_\star\frac{dI_\star}{dt}\right]
\label{blue:amdisc}
\end{eqnarray}

This expression indicates that the rate of spin down of the star is
independent of the mass of the disc. Once we have a substantial disc
we will get a braking torque on the star and it will always lead to
the same reduction in the angular velocity of the star. This is in
agreement with the results of \citet{AC96}.

If the material surrounding the merged object that was discussed in
Section~\ref{blue:ev} does collapse down to form a disc, then it would
have the same effects as the disc that has been postulated in this
section.

\subsubsection{Locking to out-flowing material}

In previous sections we have explicitly stated that we do not have
large convection zones on the surface of any of the collisional
products. This has thus apparently ruled out a torque associated with
stellar wind braking.\footnote{
In a normal main-sequence star, if we are to lose angular momentum
through a stellar wind, we require a deep surface convection
zone. This is because angular momentum transport processes within the
star are very slow. We rely on dynamical motions in the convection
zone to speed up the transport of angular momentum around the star.
However, the magnetic field of these merged objects is likely to be
very high because of their rapid rotation (see
Eqn.~\ref{blue:bfield}). Thus, it is likely that the star and
field are tightly locked.
Hence,
as a torque is applied to the magnetic field, the whole star suffers
the same torque and it is spun down.}
However, we know that the collisional product will
contract and spin up. This will eject material from the stars surface,
whilst not strictly a wind in the conventional sense, there is no
apparent reason why this material cannot become locked to the magnetic
field as it moves away from the star and thus produce a braking torque.

To examine how this torque will break the rotation of the star, we
first start off with the mass continuity equation:

\begin{eqnarray}
\dot{m}=4\pi R^2\rho v
\label{blue:mdot}
\end{eqnarray}

\noindent where $\dot{m}$ is the mass loss from the star, or
equivalently the mass flux of the wind-like material across a
surface, $v$ is the wind speed (which will be some multiple of the
escape velocity of the star) and $\rho$ is the density of the wind-like
material. 

We now look for pressure balance between the gas pressure and the
magnetic pressure. This will give us the Alfv\'{e}n radius for the
star. 

\begin{eqnarray}
\rho_{\rm ra}v_{\rm ra}^2=\frac{B_{\rm r}^2(R_{\rm ra})}{8\pi}
\label{blue:ra}
\end{eqnarray}

If we look at an open field line configuration, we may replace $B_r$
with:

\begin{eqnarray}
B_{\rm r}=B_o\left(\frac{R_\star}{R}\right)^2
\label{blue:mag}
\end{eqnarray}

\noindent If we now substitute Eqn.s~\ref{blue:mdot}
and~\ref{blue:mag} into Eqn.~\ref{blue:ra} we get:

\begin{eqnarray}
R_{\rm ra}=\frac{B_o R_\star^{2}}{\sqrt{2 \dot{m} v}}
\end{eqnarray}

\noindent then simply setting the velocity of the wind to be the
escape velocity from the surface of the star the Alfv\'{e}n radius may
be expressed as:

\begin{eqnarray}
R_{\rm ra}=B_o R_\star^{\frac{9}{4}} \dot{m}^{-\frac{1}{2}}
m_\star^{-\frac{1}{4}} \frac{1}{\left(8G\right)^{\frac{1}{4}}}
\label{blue:raf}
\end{eqnarray}

If we the equate the angular momentum loss from the star to be the
same as the angular momentum carried away by the wind once it has
reached the Alfv\'{e}n radius, one can say:

\begin{eqnarray}
\frac{dJ}{dt}=\dot{m}\Omega_\star R_{\rm ra}^2
\end{eqnarray}

\noindent then once we substitute Eqn.~\ref{blue:raf} into the
above we get:

\begin{eqnarray}
\frac{dJ}{dt}=B_o^2 R_\star^{\frac{9}{2}} m_\star^{-\frac{1}{2}}
\Omega_\star \frac{1}{\left(8G\right)^{\frac{1}{2}}}
\end{eqnarray}

This agrees with the results of \citet{V84}.

Then by using Eqn.~\ref{blue:domega} we can get the spin down rate
for the star:

\begin{eqnarray}
\frac{d\Omega_\star}{dt}=\frac{1}{I_\star}\left[B_o^2
R_\star^{\frac{9}{2}} m_\star^{-\frac{1}{2}} 
\Omega_\star \frac{1}{\left(8G\right)^{\frac{1}{2}}}
-\Omega_\star\frac{dI_\star}{dt}\right]
\label{blue:amwind}  
\end{eqnarray}

\begin{figure}
\includegraphics[clip,width=0.95\linewidth]{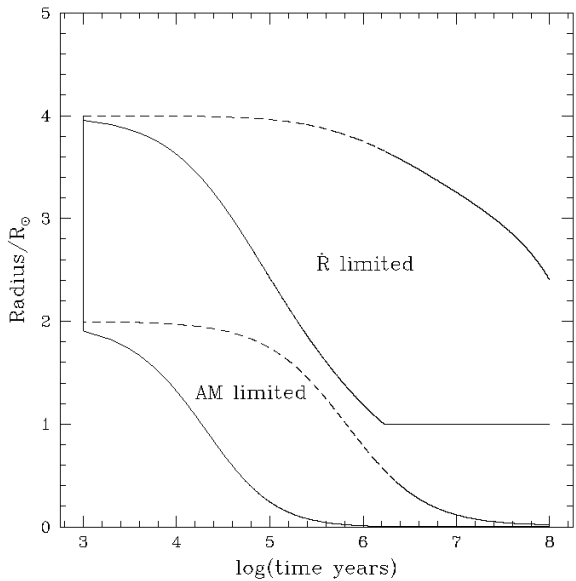}
\caption{Evolution of the radius of the blue straggler as a function
of time. The first two lines (labelled $(\dot{R})$) represent
 how the radius of the blue straggler would
evolve if it were solely dependent on the angular momentum losses
within the star. The solid line represents angular momentum losses due
to wind braking, whilst the dashed line represents the effects of
having a disc. In reality the star will shrink by more than this, as
its radius is also dependent on the cooling of the star. The point
where it can shrink to is limited by its angular momentum. The lower
two lines on the plot show the smallest radius that the star can
shrink to with the angular momentum that it has at a particular time
step. The evolution of a blue straggler will lie somewhere in between
the two encompassing lines.} 
\label{blue:radevol}
\end{figure}

One can make an estimate on the likely evolution of the radius of the
blue straggler by using this torque. As we know, the torque will slow
the star down and this will allow the star to contract to smaller and
smaller radii until an equilibrium is reached. If we assume that the
star is initially rotating at some fraction of its maximum rotation
rate:

\begin{eqnarray}
\Omega_\star=f\sqrt{\frac{GM_\star}{R_\star^3}}
\end{eqnarray}

\noindent then we can calculate the minimum size that the star can
have as a function of time. This is plotted in
Fig.~\ref{blue:radevol}. For this figure and the following, we are
looking at a representative collision product with a mass of 1 \msun,
an initial radius of 4 \rsun, and an initial rotation rate where $f$ =
0.707. We also assumed a magnetic field that had an initial value of
200 Gauss and varied with time as :

\begin{eqnarray}
B=B_o\left(\frac{P_o}{P}\right)^2
\label{blue:bfield}
\end{eqnarray}

\begin{figure}
\includegraphics[clip,width=0.95\linewidth]{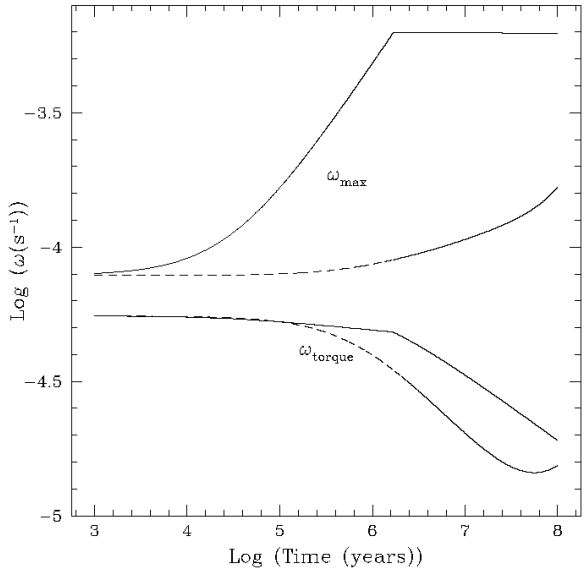}
\caption{The spin evolution of one of the blue straggler stars. The
solid lines represents the effects of angular momentum loss through a
wind, whilst the dashed lines represents angular momentum loss through
locking to a disc. The top two lines represent the maximum $\omega$
that the model could have at a particular time if its radius was
dominated by angular momentum losses (\ie it followed the upper lines
in Fig.~\ref{blue:radevol}). The lower two lines represent the
$\omega$ calculated for the models using Eqns.~\ref{blue:amdisc}
and~\ref{blue:amwind}. As can be seen, both forms of angular momentum loss
have very similar effects on the spin of the spin of the star, and
rapidly take it below its critical velocity.}
\label{blue:omega}
\end{figure}

\noindent (following \citet{AC96}), and a mass loss rate of
10$^{-10}$\msun/yr, although the form of the graph is actually largely
independent of $\dot{m}$.  In the plot we have shown two sets of
lines. The uppermost set represent the evolution of radius if it were
solely dependent on the change in angular momentum of the star. The
lower set represent the minimum radius that the star can contract to
with the angular momentum that it has at a particular time step. As
can be seen, we quickly loose sufficient angular velocity to allow the
blue straggler to contract down via either angular momentum loss
method. Once the star has contracted to its desired radius, the loss
of angular momentum will continue, but now the star will maintain its
radius and spin more slowly. In Fig.~\ref{blue:omega} we show the spin
evolution of the same blue straggler. The lower set of lines represent
the $\omega$ that the star has because of the torque applied to the
star. The upper set of lines represents the maximum $\omega$ that a
model can have at a particular time step if its radius is dominated by
the angular momentum losses, \ie~the radius of the star follows the
upper-most set of lines in Fig.~\ref{blue:radevol}. As can be seen,
both forms of angular momentum loss have similar effects on the star,
spinning it down to lower rotational velocities.

\subsection{A detailed evolutionary calculation for an off-axis collision product with angular momentum loss}

We chose one collision product to model in detail using a stellar
evolution code, including the effects of rotation and the angular
momentum loss methods described above. Angular momentum transport
inside the star and the thermal contraction of the star are additional
mechanisms which could substantially impact the evolution of the blue
straggler. In this section, we look at those effects for one
representative case, the M66Q case ($M_1=0.6$ \msun, $M_2=0.6$ \msun,
$r_p=0.25 R_1+R_2$).

\subsubsection{The Stellar Evolution Code}

Our stellar evolution calculations are performed with YREC. YREC is a
one-dimensional code, in which the star is divided into shells along
surfaces of constant gravitational plus rotational potential. The code
solves the equations of stellar structure with the Henyey technique,
and follows the rotational evolution with the formalism of
\cite{ES76}.  \cite{GDPK92} give a detailed description of the
physics implemented in the evolution code. We used the same opacities,
equation of state and model atmospheres as in \citet{SLBDRS97}. The free
parameters in the code (the mixing length and parameters that set the
efficiency of angular momentum transport and rotational chemical
mixing) are set by calibrating a solar mass and solar metallicity
model to the sun.

Rotation is treated by evaluating physical quantities on equipotential
surfaces rather than the spherical surfaces usually used in stellar
models. The hydrostatic equilibrium and radiative transport equations
contain terms that account for the lack of spherical symmetry in the
rotating star. A number of rotational instabilities that transport
angular momentum and material within the star are followed, including
dynamical shear \citep{PKSD89}, meridional circulation \citep{vZ24},
secular shear \citep{Z74}, and the Goldreich-Schubert-Fricke
instability \citep{GS67,F68}. Angular momentum transport and the
associated chemical mixing are treated as diffusion processes, with
diffusion coefficients that account for each active mechanism within
unstable regions of the star.  The diffusion coefficients are
proportional to circulation velocities, which have been estimated by
\citet{ES78}.  In addition to the internal rearrangement of angular
momentum, angular momentum also can be drained from an outer
convection zone through a magnetic wind, using the formalism given by
\citet{CDP95}. For a detailed description of the implementation of
rotation in YREC, see \citet{PKSD89}.

\subsubsection{Conversion to YREC Format}

The results of the three-dimensional SPH simulations were converted to
one-dimensional models suitable to use as starting models for
YREC. The specific entropy of the particles (defined as $A =
P/\rho^{\gamma})$, their chemical abundances and their specific angular
momentum were averaged over surfaces of constant effective potential
(gravitational plus rotational), and binned into approximately 100
bins. With the entropy and angular momentum profiles given, the
structure of the rotating remnant is uniquely determined in the
formalism of \citet{ES76} by integrating the general
form of the equation of hydrostatic equilibrium. To do so, we
implement an iterative process in which initial guesses for the
central pressure and angular velocity are refined upon until a
self-consistent YREC model is converged upon. For more details of the
conversion between SPH results and stellar evolution models, see \citet{SLBDRS97, SFLRW01}.

\subsubsection{Early Evolution and Treatment of Angular Momentum Loss}

Immediately after the collision, the collision products have large
radii and are quite far out of thermal equilibrium. Their early
evolution is dominated by a contraction back to the main
sequence. Since they are not homogeneous pre-main sequence stars,
however, this contraction produces a track (for non-rotating stars)
which moves down the middle of the HR diagram rather than on the cool
Hayashi track. Rotating stars have an additional complication. These
stars are rotating quite rapidly (surface rotation velocities of
$\sim$ 300 - 700 km s$^{-1}$, depending on the star), although less than
their surface break-up velocities at this point. As
the star contracts, its surface rotation rate increases due to angular
momentum conservation. The radius shrinks as well, which increases the
break-up velocity, but unfortunately for the star, the angular
momentum wins. Shortly after the collision, the star is rotating
faster than its break-up velocity, and mass will be thrown off of the
star. This should lead to a self-regulating process where the star
continues to rotate at its break-up velocity, shedding mass as
required to maintain that rate.

Standard stellar evolution codes do not usually allow the star to lose
mass. We have implemented a mass loss routine in YREC which removes
mass from a surface convection zone if one exists, or from the entire
star if one does not. All our blue straggler candidates fall in the
latter category. We calculate what fraction of the star is rotating at
higher than its local break-up velocity at a given time, and then (to
ensure numerical stability of our models) remove that fraction from
each of the shells in the star. We assume that the mass loss occurs
instantaneously, so that the changes to radius, temperature, density,
and pressure of each shell will occur during the evolutionary time
step following the mass loss. We need to remove the angular momentum
that leaves the star with the mass. We do this by assuming that the
profile of angular momentum as a function of mass fraction in the star
remains the same as before, but is truncated at the new outer edge of
the star. The same procedure is used to determine the new composition
of each shell, so that we are not artificially removing helium
directly from the core of the star. This new configuration is allowed
to evolve for another timestep (typically on order 100 years at this
early stage of the evolution) and then the process is repeated.

It is plausible to assume that the mass that is lost from the star can
do one of two things: either it is lost from the star to infinity
immediately, or it remains in a disc around the star for some time. If
the star has a magnetic field, the star can become locked to the disc
so that its surface rotation rate is determined by the rotation rate
of the disc. This disc-locking is assumed to be important in the
rotational evolution of pre-main sequence stars
\citep*{K91,SPT00,BSP01}, and may well be applicable to blue
stragglers (although see \citet{MP04}). We have added a module to the
stellar evolution code which allows this disc-locking to be turned on
for a given amount of time (which is a user-set parameter). The
surface rotation rate of the star is fixed to be the surface rotation
rate immediately after the mass was lost. If this module is turned on,
the disc-locking turns on after the star has lost at least 0.1 \msun.

\subsubsection{Results}

We ran two evolutionary calculations. The collision product
was evolved until it threw off the minimum mass required for
disc-locking to turn on. Then, we either allowed the star to continue
evolving without the disc-locking, or locked the star to the disc for
5 million years. The lifetimes for pre-main sequence discs are
thought to range between 0 and 10 million years \citep{SSECS89}. 

\begin{figure}
\includegraphics[clip,width=0.95\linewidth]{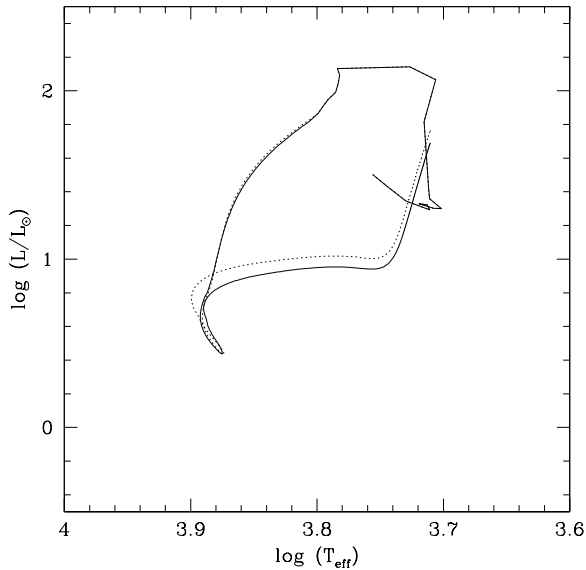}
\caption{Evolutionary tracks in the Hertzsprung-Russell diagram for the 
M66Q collision, with (solid line) and without (dotted line) disc-locking.  
\label{tracks}}
\end{figure}

The evolutionary tracks of the collision products with disc-locking
(solid lines) and without (dotted lines) are presented in Fig.
\ref{tracks}. Initially the star lost 0.10 \msun during its
super-critical rotation period. The final angular momentum after the
mass loss was $9.0 \times 10^{49}$ g cm$^2$ s$^{-1}$, and after the
star had been locked to a disc for 5 Myr, its total angular momentum
was $1.5 \times 10^{49}$ g cm$^2$ s$^{-1}$ (i.e. the star loses $\sim
80$\% of its angular momentum during the 5 Myr disk-locking
period). Note that the stars lose$\sim$ 90\% of their initial angular
momentum during the mass loss phase. The material at the outer edge of
the star has very large specific angular momentum, so a little bit of
mass can remove a lot of angular momentum.

While both stars initially begin their post-collision evolution in the
lower right portion of this diagram, the star that was locked to a
disc follows a fairly normal evolutionary track.  This star is still
significantly brighter than a normal star of the same mass and shows
some evidence of rotational mixing and modification of evolutionary
track and lifetime. The star that was not locked to a disc (dotted
line) mixes more hydrogen to the core and helium to the surface. Its
evolutionary track is slightly bluer and brighter, and its main
sequence lifetime is slightly longer than the disk-locked star.

\begin{figure}
\includegraphics[clip,width=0.95\linewidth]{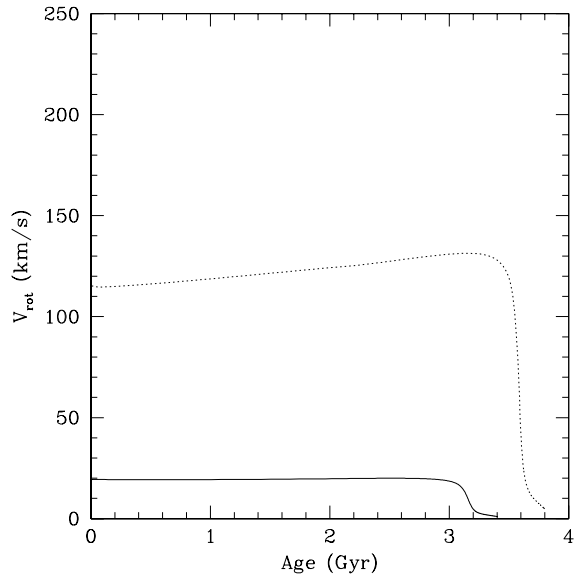}
\caption{Rotation rate as a function of time for the M66Q collision, 
with (solid line) and without (dotted line) disc-locking. The shaded
region shows the current range of detected rotation rates for blue
stragglers in globular clusters: 155 $\pm$ 55 km s$^{-1}$ (BSS-19 in 47 Tuc,
\citet{SSL97}) and 200 $\pm$ 50 km s$^{-1}$ (BSS-17 in M3,
\citet{DLOZS04}). \label{rotation}}
\end{figure}

In Fig. \ref{rotation}, we plot the surface rotation velocity in km
s$^{-1}$ as a function of time for the M66Q collision, with the solid
line showing the star which was locked to a disc, and the dotted line
for the star which was allowed to evolve freely after it lost
mass. The difference in surface rotation rates is quite apparent, and
directly reflects the amount of angular momentum which was carried
away by the disc. Rotation rates of 125 km/s are quite fast for stars
of 1 \msun (our Sun rotates at $\sim$ 2 km s$^{-1}$), but are not
completely unreasonable given the observed rotation rate of a few blue
stragglers in a globular cluster. The ordinate of this graph gives the
range of observed rotation rates \citep{SSL97,DLOZS04}.

\begin{figure}
\includegraphics[clip,width=0.95\linewidth]{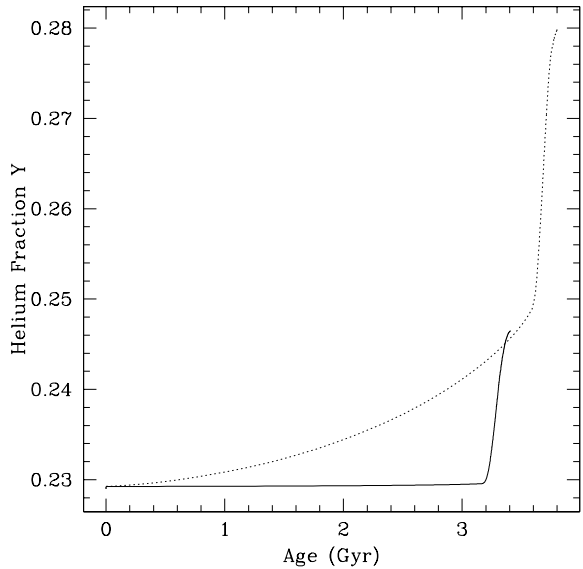}
\caption{Helium abundance as a function of time for the M66Q collision, 
with (solid line) and without (dotted line)
disc-locking. \label{helium}}
\end{figure}

\begin{figure}
\includegraphics[clip,width=0.95\linewidth]{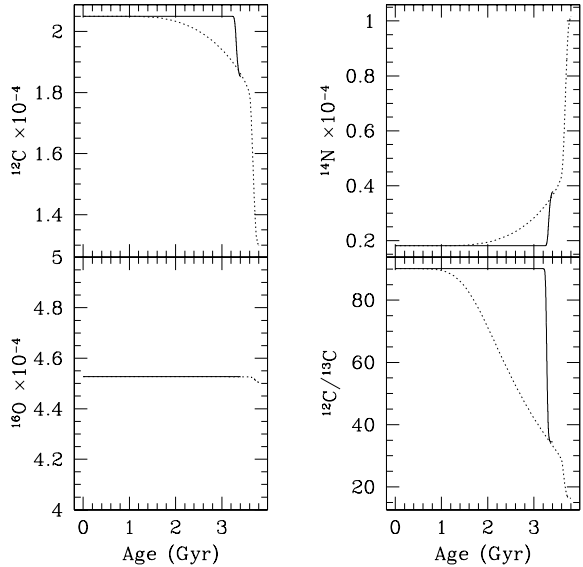}
\caption{Surface composition as a function of time for the M66Q collision,
 with (solid line) and without (dotted line) disc-locking. \label{composition}}
\end{figure}

An interesting effect of rotation in stars is the amount of
rotationally-induced mixing that can occur. Since rotating stars are
subject to a variety of thermodynamic and hydrodynamic instabilities,
more mixing of material can occur than that caused by convection. This
mixing, particularly of hydrogen to the core and helium to the
surface, is responsible for the extended lifetime of the rotating
stars, and is shown in Fig. \ref{helium}. This plot of surface helium
abundance as a function of time for the M66Q collision shows that
neither star ever becomes fully mixed, but that the non-disk-locked
star does show some rotationally-induced mixing over its main sequence
lifetime.

Other elements also trace this rotation mixing. The CNO elements, for
example, show strong changes are CNO processed material is dredged up
from the nuclear burning regions to the surface. These changes are
shown in Fig. \ref{composition}. As with the helium abundance, the
star which was not locked to a disc shows a much stronger signature of
rotational mixing than the star with a disc. The clearest evidence of
this mixing shows up in the $^{12}C/^{13}C$ ratio, which deviates from
the canonical main sequence value of 90 fairly early in its main
sequence lifetime.  This ratio may be a crucial diagnostic for
determining the rotational histories of blue stragglers.

The choice of disc lifetime, at 5 Myr, was based on our understanding
of discs around pre-main sequence stars and was somewhat arbitrary. We
ran an identical simulation with a 10 Myr disc to test the extreme of
the plausible range. The majority of the angular momentum was lost
from the star in the first few Myr, and so extending the disc lifetime
did not make a substantial difference to the star's evolution. The
total angular momentum of the blue straggler was 1.493 $\times 10^{49}$
g cm$^2$ s$^{-1}$ after 10 Myr (compared to 1.498 $\times 10^{49}$ g
cm$^2$ s$^{-1}$) and at an age of 2 Gyr, the star with the 10 Myr disc
had $v \sin i = $ 19.7 km s$^{-1}$ compared to 19.8 km s$^{-1}$ for the
star with a 5 Myr disc. Therefore, we conclude that the disk needs to
survive for only a few Myr in order to slow the star down substantially.

An additional level of arbitrariness was setting the mass of the disk
beyond which disk locking occurs. To test this, we locked our star to
its disk as soon as it lost any mass at all. Under this scenario, the
star's surface rotation rate was fixed at $\omega = 5.9 e-5$ rad
s$^{-1}$ for 5 Myr starting almost immediately after the
collision. This produced a significantly different evolution than that
shown in figure \ref{tracks}. The star loses a total of 0.42 \msun
rather than 0.1 \msun. This is understandable -- if we wait to lock
the star's rotation rate, then the star has more time to contract
towards the main sequence. This thermal contraction rate is largest
early on. As the star contracts, it is more likely to exceed its
maximum rotation speed for a given $\omega$, and so the star ends up
losing much more mass. Interestingly, the final rotation rate of this
new low-mass star is approximately the same as the star which was
locked to a disk after it lost 0.1 \msun: about 20 km/s. Therefore,
the disk mass at which locking can occur could change the total mass
of the blue straggler, but apparently has less impact on its
rotational properties. We argue that a disk mass of 0.1 \msun is
reasonable based on our understanding of disk-locking in pre-main
sequence stars.

We also tested our assumption that the star is locked at the surface
rotation rate immediately after the mass is lost. The star is rotating
quite rapidly at this point, so this assumption can be taken to be a
reasonable upper limit to the rotation rate of the star. If the star
manages to lose more mass than we assumed, or if it is locked to the
disk at a lower surface rotation rate, than the blue straggler will be
rotating even slower than we have shown in the previous
calculations. We ran the same simulations as discussed above, except
that at the time when the star was locked to its disk, we locked it at
one-fifth of the current surface rotation rate. The factor of 5 was
chosen to represent a reasonable reduction in rotation rate. We ran
the simulations with both a 5 Myr and a 10 Myr disk. As expected, the
total angular momentum of the system after the disk-locking episode
was approximately a factor of 5 lower than in the original
calculations (5.2 times lower for the 5 Myr disk and 4.97 times for
the 10 Myr disk). The surface rotation rates were lower by a factor of
$\sim$ 4.

\section{Discussion and Conclusions}

We used SPH to simulate stellar collisions relevant for the formation
of blue stragglers in globular clusters. Our results were the same as
seen in other papers, with the possible exception of the lack of
discs. We believe that the higher resolution of our calculations and
the use of realistic stellar models (rather than polytropes) is the
root of the discrepancy between our simulations and those of
\citet{BH87}.  It should also be noted that the discs observed around blue
stragglers by \citet{DLOZS04} have masses much lower than the particle
masses in our SPH simulations, and so the simulations and observations
are not in disagreement.

We have shown that magnetic locking to either a disc or to outflowing
material would produce sufficient angular momentum loss to avoid
complete break-up of the blue straggler.  We have shown that if the
star is locked to a disc of material via a magnetic field, enough
angular momentum is lost for the star to evolve along a basically
normal, but brighter, evolutionary track. If, on the other hand, the
star is allowed to lose mass but is not locked to a disc, it continues
to evolve at a significant fraction of its surface break-up
velocity. Substantial amounts of hydrogen and helium are mixed
throughout the star via the rotational instabilities, and the star
follows a path in the colour-magnitude diagram which runs up the
zero-age main sequence. The star becomes very hot and very blue, and
has a greatly extended main sequence lifetime.

We see some very bright blue stragglers in some clusters (M3, NGC
6752, NGC 6397). In other clusters, the blue stragglers are not
particularly bright, nor are they constrained to lie on the zero-age
main sequence \citep{FSRPB03}. Instead, they are all to the red of the
ZAMS. Therefore, the highly rotational models (i.e. those without disc
locking) do not match most of the observations well. Either most blue
stragglers are locked to a disc and therefore evolve more like
non-rotating stars, or stellar collisions are not the dominant
creation mechanism for these stars. It is possible that the brightest,
bluest blue stragglers in clusters are indeed rotating rapidly, and
that that is the cause of their position in the colour-magnitude
diagram. However, the stars in question are not as bright or as blue
as the predictions from the no-disc models, so we conclude that the
disc-locking mechanism, or some similar angular momentum loss
mechanism, must be at work during the blue straggler formation
process.

Is it reasonable to assume that the collision product can be locked to
a disc? Low mass stars all have magnetic fields, so if the field can
be maintained and remains coherent during the stellar collision, then
the product should also have a field of the appropriate strength. It
is difficult to understand how the magnetic field could be completely
removed from the stellar material during the collision, although the
stirring up of stellar material may indeed weaken it. More work is
needed to determine if the magnetic field of the collision product
will be coherent enough at the end of the collision to provide the
locking required.

There are a number of observational tests which will shed some direct
light on the nature of blue stragglers as rotators. The most obvious
is measurements of rotation rates of blue stragglers from spectra, as
was done for BSS 19 in 47 Tuc \citep{SSL97}. HST STIS spectra have
been taking for blue stragglers in a number of clusters, but the
reductions have not yet been published. The early indications are that
the blue stragglers are rotating very slowly, less than $v\sin i = 50
$km s$^{-1}$. (Shara, private communication, 2003). These spectra will also
be incredibly useful for determining surface gravity values for blue
stragglers, which lead to mass determinations. A comparison of derived
mass with location in the colour-magnitude diagram will allow us to
determine if blue stragglers have evolutionary tracks like normal main
sequence stars, or if they are more like the rotational evolutionary
tracks presented in this paper. Particularly for the brighter blue
stragglers, the masses will allow us to distinguish between
rotationally-modified evolutionary tracks and stars with masses more
than twice the turnoff mass.  An additional spectral diagnostic for
rotation is the abundances of a variety of elements, particularly the
CNO elements and helium. The rotationally-induced mixing signatures
can be particularly strong and fairly unambiguous, although the CNO
abundances at the surface of a binary merger product are yet to be
determined. Since collisional blue stragglers should live at the
centres of dense clusters, observing them spectroscopically will be
quite challenging and points to the use of the Hubble Space Telescope
and ground-based adaptive optics systems. However, there are now more
and more telescopes equipped to perform this kind of work, which would
significantly reduce the level of uncertainty about these interesting
and useful members of globular clusters.

\section{Acknowledgments}
AS is supported by NSERC. TAD acknowledges a PPARC studentship. MBD is
supported by a research fellowship from the Swedish Royal Academy of
Sciences and PPARC funding for theoretical astronomy at the University
of Leicester.  The computations reported here were performed using the
UK Astrophysics Fluids Facility (UKAFF) and the SHARCNet facilities at
McMaster University.

\end{document}